\documentclass[epjST]{svjour}
\usepackage{graphics}
\usepackage{amsmath}
\usepackage{epstopdf}

\def\GG{\textbf{G}}
\begin{document}
\title{Mesoscopic mean-field theory for spin-boson chains in quantum optical systems}
\author{Pedro Nevado\inst{1}\fnmsep\thanks{\email{pnevado@estumail.ucm.es}}
\and
Diego Porras\inst{1}\fnmsep\thanks{\email{diego.porras@fis.ucm.es}}}
\institute{\inst{1} Departamento de F\'isica Te\'orica I, Universidad Complutense, 28040 Madrid, Spain}
\abstract{
We present a theoretical description of a system of many spins strongly coupled to a bosonic chain. We rely on the use of a spin-wave theory describing the Gaussian fluctuations around the mean-field solution, and focus on spin-boson chains arising as a generalization of the Dicke Hamiltonian. 
Our model is motivated by experimental setups such as trapped ions, or atoms/qubits coupled to cavity arrays. 
This situation corresponds to the cooperative (E$\otimes$$\beta$) Jahn-Teller distortion studied in solid-state physics. 
However, the ability to tune the parameters of the model in quantum optical setups opens up a variety of novel intriguing situations. 
The main focus of this paper is to review the spin-wave theoretical description of this problem as well as to test the validity of mean-field theory. Our main result is that deviations from mean-field effects are determined by the interplay between magnetic order and mesoscopic cooperativity effects, being the latter strongly size-dependent.}
\maketitle
\section{Introduction}
\label{intro}

Experimental progress in quantum optical setups has opened up new research areas where the controllability of those experimental systems meets the complexity of quantum many body physics. For example, trapped ions and ultracold atoms can now be controlled to the extent in which they emulate the physics of Condensed Matter systems in a way that may lead to the implementation of quantum simulators of many-body models 
\cite{Cirac12natphys,Schneider12rpp}. The latter are devices in which quantum states may be prepared and measured and interactions may be tuned, so as to mimic complex dynamics with a practical and scientific interest in material science. The research field of Analogical Quantum Simulation has emerged to exploit this idea.

Remarkably the physical elements of many of those systems, particularly trapped ions and atoms interacting with photons, can be understood by means of spin-boson models. 
Effective spins are implemented by two-level systems which can be either atomic transitions of trapped ions \cite{Leibfried03rmp,Porras04aprl}
or isolated energy levels in solid-state qubits \cite{Hartmann06natphys,Schoelkopf08nat}. 
The bosonic degree of freedom is provided by vibrations in ion Coulomb crystals 
\cite{Porras04bprl} or photons in optical or microwave cavities. 
Bosonic modes may be collective excitations either because of the vibrational couplings in the case of ions, or because of the collective nature of photonic bands in cavities. Finally, a number of different spin-boson couplings may be implemented in those systems, either by laser-induced forces in the case ions, or by the light-matter coupling between qubits and cavities. To focus our discussion, let us define 
$\sigma_{j}^z = |1\rangle_j \langle 1|-|0\rangle_j \langle 0|$,with $|0\rangle_j, |1\rangle_j$ internal energy levels of an atom (qubit), and \smash{$a_j$, $a_j^\dagger$} annihilation, creation operators for a bosonic mode at site $j$. A natural coupling in quantum optical systems is
\begin{equation}
H_{\rm I} = g \sum_j \sigma^z_j \left( a_j + a^\dagger_j \right).
\label{intro}
\end{equation}
If an additional process couples different bosonic modes, qubit-boson couplings provide us with a mechanism to couple different qubits. This idea is the basis for quantum gate designs \cite{Cirac00nat,Schoelkopf08nat,Peropadre10prl}. 
In the field of Analogical Quantum Simulation, the use of couplings like (\ref{intro}) has been mainly an instrument to get effective models, like quantum magnetic Hamiltonians, in which tracing out bosons results in effective interactions of the form $\sum_{j,k} J_{j,k} \sigma^z_j \sigma^z_k$ 
\cite{Porras04aprl,Deng05pra,Friedenauer08natphys,Bermudez09pra,Islam11natcomm,Britton12nat}.
Also, strongly coupled spin-boson chains have been considered in a few previous works, for example in the context of disordered systems \cite{Bermudez10njp} and quantum dissipation involving a single spin interacting with a bosonic bath \cite{Porras08pra}.

Quite remarkably, however, interaction (\ref{intro}) poses an outstanding many-body model defined on a quantum lattice, with a quantum phase diagram which is intrinsically interesting. It belongs to a family of many-body models previously considered in the context of solid-state physics, in particular, in the description of the cooperative Jahn-Teller 
E$\otimes$$\beta$ distortion \cite{Englman.book}. 
The latter is a many-body effect which appears in solids, where two-level systems correspond to orbital electronic levels localized in atoms, which interact with the lattice vibrations. This effect has received attention for its connections with high-Tc superconductivity and colossal magnetoresistance \cite{Millis96prb,Millis96prl,Tokura00sci}. 
Very recently it was shown that a generalization of interaction (\ref{intro}) can be implemented in a quite natural way in a linear ion crystal placed in a magnetic field gradient \cite{Porras12bprl,Ivanov12arX}.
The prospects for experimental investigation of the cooperative Jahn-Teller effect would be very interesting, since they would allow us to observe novel quantum structural phase transitions in clean controllable systems.  Solid-state systems pose many limitations to observe quantum effects in cooperative Jahn-Teller models, since experiments are typically performed at high temperatures, and also because Jahn-Teller couplings are masked by magnetic or electronic dynamics. A variety of related models are currently under theoretical investigation by several groups including 
the single particle case \cite{Casanova10prl},
Jaynes-Cummings-Hubbard models \cite{Leib10njp,Tureci12prl}, structural phase transitions in quantum potentials \cite{Cormick12prl} and spin-Peierls transitions \cite{Bermudez12prl}.

This paper is dedicated to the spin-wave theory of cooperative 
(E$\otimes$$\beta$) Jahn-Teller models arising from longitudinal 
$\sigma^z$ couplings of the form (\ref{intro}). 
A natural starting point is mean-field theory together with a spin-wave description of Gaussian fluctuations \cite{Auerbach}. The main goal of this paper is to test the validity of mean-field theory by calculating the intensity of quantum fluctuations. The outline of the paper is the following. In section 2 we introduce the Hamiltonian that we are considering for cooperative Jahn-Teller systems, and present some qualitative considerations arising from the comparison with the Dicke model. In section 3 we review the mean-field in this model. In section 4 we present a calculation of the Gaussian fluctuations by means of the definition of Holstein-Primakoff bosons. 
In section 5 we study in detail the case of Periodic Boundary Conditions and check the validity of the mean-field approximation in a region that depends both on the magnetic order of the system, as well as on mesoscopic cooperative effects as a function of the number of particles. Finally in section 6 we present our conclusions.
\section{Cooperative Jahn-Teller systems}
\label{sec:cJT_model}

We consider a chain of $N$ spins coupled to $N$ bosonic modes. In addition, there are couplings between different bosonic modes as well as a magnetic field acting on the spins. This leads to a cooperative Jahn-Teller Hamiltonian of the form,
\begin{eqnarray}
H_{\rm JT} &=& H_{\mathrm{s}}+H_{\mathrm{b}}+H_{\mathrm{sb}},
\nonumber \\
H_{\mathrm{s}} &=& \sum_{j}\frac{\Omega }{2}\sigma _{j}^{x}, \ \ \ \ 
H_{\mathrm{b}} =\sum_{j}\omega _{j}a_{j}^{\dagger
}a_{j}+\sum_{j,l}t_{j,l}a_{j}^{\dagger }a_{l}, \ \ \ \ 
H_{\mathrm{sb}} = g \sum_{j}\sigma _{j}^{z}(a_{j}+a_{j}^{\dagger }),
\label{Hamiltonian}
\end{eqnarray}
where $a_j$, $a_j^\dagger$ are annihilation/creation operators of bosons that are localized near the effective spin. 
The physical meaning of those bosonic modes can be either vibrations of a trapped ion, or photons confined in a cavity. 
$\Omega$ is a transverse magnetic field, and $\omega_j$ and $t_{j,l}$ are boson local energies and hopping amplitudes; the latter determine the boson dispersion. 
% We only consider short-range couplings between bosonic modes,
There are several experimentally relevant cases. However, in this work we focus on nearest-neighbours boson tunneling, 
\begin{equation}
t_{j,l} = -t(\delta _{j,l+1}+\delta _{j,l-1}),\,t > 0 .
\label{tunn}
\end{equation}
The latter describes coupled cavities,
 and to a fair approximation also the radial phonons of an ion Coulomb chain \cite{Deng08pra}.
We diagonalize the bosonic Hamiltonian by transforming to a collective mode basis,
$a_j = \sum_n M_{j,n} \bar{a}_n$, where $M_{j,n}$ is the collective mode amplitude at site $j$. The relation
\begin{equation}
\bar{\omega}_{n} \delta_{n,m} =
\sum_{j,l}
M_{j,n}^*(t_{j,l} + \omega _{j} \delta _{j,l}) M_{l,m}
\end{equation}
determines the collective mode energies, $\bar{\omega}_n$, such that 
$H_{\rm b} = \sum_n \bar{\omega}_n \bar{a}^{\dagger}_n \bar{a}_n$. Finally, the spin-boson coupling in the collective modes basis reads
\begin{equation}
H_{\mathrm{sb}}
=
g\sum_{j,n} \sigma _{j}^{z} 
\left(  M_{j,n} \bar{a}_{n} + M_{j,n}^* \bar{a}_{n}^{\dagger} \right) .
\end{equation}
%
% Later on we will focus on Periodic Boundary Conditions (P.B.C.), such that 
The problem posed by $H_{\rm JT}$ resembles the Ising quantum magnet 
in a transverse field. 
For large $\Omega$ the ground state is a product state of the boson vacuum and spins pointing in $x$. In the case of large $g$ magnetic order is in the $z$-direction.
To show this result, let us see how to get effective spin-spin Ising interactions 
\cite{Porras04aprl,Deng05pra}. 
Consider the canonical transformation $U=e^{-S}$ with  
\begin{equation}
S = \sum_{j,n} (g/\bar{\omega} _{n}) \sigma _{j}^{z}
\left(M_{j,n} \bar{a}_n - M_{j,n}^*\bar{ a}_{n}^{\dagger}  \right),
\end{equation}
representing a spin-dependent boson displacement. We apply $U$ to the $\Omega = 0$ case,
\begin{equation}
e^{-S} (H_{\rm b} + H_{\rm sb}) e^{S} = 
\bar{H}_{\rm b}+\bar{H}_{\rm sb}=\sum_{n} \bar{\omega}_n \bar{a}^{\dagger}_n \bar{a}_n -\sum_{j,l,n} \frac{g^2}{\bar{\omega}_{n}}M_{j,n}^* \sigma^{z}_{j} \sigma^{z}_{l}M_{l,n},
\end{equation}
which shows that the $\Omega = 0$ Jahn-Teller model is equivalent to an Ising model with interaction strength, 
\begin{equation}
J_{j,l}=-g^{2}\sum_{n}M_{j,n}^*\frac{1}{\bar{\omega} _{n}}M_{l,n}.
\label{interaction}
\end{equation}
The former argument is equivalent to the adiabatic elimination of bosons \cite{Deng05pra}. Adding the transverse field term, we find 
\begin{equation}
e^{-S} H_{\rm s} e^{S} = 
\bar{H}_{\mathrm{s}}=\frac{\Omega }{2}\sum_{j}\left( \sigma
_{j}^{+}e^{-2\sum_{n}M_{j,n}\frac{g}{\bar{\omega} _{n}}\left( \bar{a}_{n}-\bar{a}_{n}^{\dagger
}\right) }+\mathrm{H.c.}\right) ,
\end{equation}
which shows that as long as $g \ll \bar{\omega}_n$, $H_{\rm s} \approx \bar{H}_{\rm s}$. Previous works exploited this idea for the quantum simulation of Ising and Heisenberg models with trapped ions \cite{Porras04aprl}. 
Here we are interested on the whole phase diagram of the model, where bosons suffer a displacement which cannot be neglected.

We aim to give a mean-field description and to test its validity. Let us first qualitatively discuss the problem. Mean-field theory may be well justified by two lines of reasoning that arise in two different fields: 
\begin{itemize}
\item {\bf Quantum magnetism.-} The limits of large $\Omega$ and large $g$ should present a well-defined magnetic order in the $x$ and $z$ directions, respectively. Mean-field theory is a good approximation there. 
\item {\bf Cooperativity effect.-} In the limit of large $t$, we expect $\bar{\omega}_n$ to form a set of energies with a large energy separation $\Delta \bar{\omega} \propto t$. In the case that $\Delta \bar{\omega} \gg g$, we may simply ignore high energy modes and keep $n=0$ only. We recover the celebrated Dicke model with infinite range interactions, for which mean-field theory is exact in the thermodynamic limit \cite{Hepp73}. However, as $N \to \infty$ we expect $\Delta \bar{\omega} \to 0$ so that this argument is only valid in a mesoscopic regime of spin-boson chains.
\end{itemize}
{\it The interplay between magnetic order and finite-size effects is a unique feature of the spin-boson model described by $H_{\rm JT}$}.

\section{Mean-field theory}
\label{sec:mf_theory}

Let us present our mean-field variational ansatz, which consists of a product state of spins in the $x$-$z$ plane, and bosons displaced in the collective mode basis \cite{Porras12bprl},
\begin{equation}
| \Psi_{\rm MF} \rangle = \bigotimes_j 
| \theta_j \rangle  \otimes 
e^{-\sum_n (\bar{\alpha}^*_n \bar{a}_n-\bar{\alpha}_n \bar{a}^\dagger_n )} | 0 \rangle_{\rm b},
\end{equation}
where $| \theta_j \rangle = 
\cos (\theta_j /2) |0 \rangle_j + \sin (\theta_j /2) | 1 \rangle_{j}$. 
The mean-field energy is
\begin{equation}
\langle \Psi_{\rm MF} | H_{\rm JT} | \Psi_{\rm MF} \rangle = 
\sum_n \bar{\omega}_n \bar{\alpha}_{n}^* \bar{\alpha}_n - 
g \sum_{j,n} \cos \theta_j (M_{j,n}^* \bar{\alpha}_{n}^{*} + M_{j,n} \bar{\alpha}_n)
+\frac{\Omega}{2} \sum_j \sin \theta_j.
\end{equation}
We minimize the energy and get a set of coupled equations for the variational parameters $\theta_j$ and $\bar{\alpha}_n$,
\begin{flalign}
&\left\{
\begin{array}{l}
\bar{\alpha}_n=\frac{g}{\bar{\omega}_n} \sum_j M_{j,n}^* \cos \theta_j,\\
\\
\sum_l J_{l,j} \cos \theta_l=- \frac{\Omega}{2} \cot \theta_j, \quad J_{l,j}=2 \sum_n \Re\left( \frac{g^2}{\bar{\omega}_n} M_{j,n}^*M_{l,n}\right).
\end{array}
\right.
\label{min_lig}
\end{flalign}
Let us consider Periodic Boundary Conditions (PBC) to simplify the discussion, together with the nearest-neighbours coupling (\ref{tunn}). 
The problem is diagonalized by plane-wave modes such that
\begin{equation}
\bar{\omega}_n = \bar{\omega}_0 + 2t (1 - \cos(2 \pi n / N)), 
\end{equation}
with $\bar{\omega}_0$ the lowest mode energy; we have assumed constant local energies in (\ref{Hamiltonian}), $\omega_j = \bar{\omega}_0 + 2 t$. 
The latter is a convenient parametrization so that the lowest collective energy mode is $\bar{\omega}_0$, independent on the value of $t$.
Collective mode amplitudes become simply 
\begin{equation}
M_{j,n} = e^{-i \frac{2 \pi n}{ N} j}/ \sqrt{N} ,
\label{pbc}
\end{equation} 
with $n = 0, \dots, N-1$. It is worth noting that the eigenfunctions are normalized and that they follow the closure relation 
$\sum_{j} e^{i\frac{2 \pi j}{N} (m-n)}=N \delta_{m,n}$.

By virtue of the PBC we can drop the index in $\theta_j,\theta_l$ in the equations (\ref{min_lig}), turning out that

\begin{flalign}
\bar{\alpha}_n&=\frac{g}{\bar{\omega}_n} \sum_{j} M_{j,n}^* \cos \theta = \frac{g}{\bar{\omega}_n} \sum_j \cos \theta \frac{1}{\sqrt{N}} e^{i \frac{2\pi n}{N} j} \notag\\
&= \frac{g}{\bar{\omega}_n} \sqrt{N} \cos \theta \delta_{n0}=
\left\{
\begin{array}{l}
\bar{\alpha}_{n \neq 0} =0, \\
\bar{\alpha}_0=\frac{g}{\bar{\omega}_0} \sqrt{N} \cos \theta,
\end{array}
\right.
\label{alpha}
\end{flalign}
and that
\begin{flalign}
\sum_l J_{l,j}&=\sum_{l} 2g^2 \Re \left( \sum_n  M_{j,n}^* \frac{1}{\bar{\omega}_n} M_{l,n} \right) = 2g^2 \Re \left(\sum_{n,l} \frac{1}{N}  e^{i\frac{2\pi n}{N}j} \frac{1}{\bar{\omega}_n} e^{-i\frac{2\pi n}{N}l} \right) \nonumber \\
&= 2g^2 \Re \left( \sum_n \frac{1}{N}  e^{i\frac{2\pi n}{N}j}  \frac{1}{\bar{\omega}_n} N \delta_{n0} \right)=\frac{2g^2}{\bar{\omega}_0}:=J.
\label{J}
\end{flalign}
Inserting (\ref{J}) in the mean-field Eqs. (\ref{min_lig}) we find the minimum at
\begin{eqnarray}
\sin(\theta) &=& -\Omega / (2 J), \ \ |\Omega|/(2 J) \leq 1,
\nonumber \\
\sin(\theta) &=& -1, \hspace{1.2cm} |\Omega|/(2 J) > 1.
\label{critical}
\end{eqnarray}
Thus, mean-field theory predicts a phase transition at a critical value
$|\Omega_{\rm c}| = 2J$. 
The behaviour of spins on both sides of the critical point can be defined by taking the two different asymptotic limits,
\begin{flalign}
&\left\{
\begin{array}{l}
\Omega \gg J \Rightarrow \cos \theta =0 \rightarrow | \theta \rangle = \frac{1}{\sqrt2}(|0\rangle \pm |1\rangle ), \\
\\
\Omega \ll J \Rightarrow \sin \theta =0 \rightarrow | \theta \rangle = \{ |0 \rangle, \pm |1\rangle \}.
\end{array}
\right.
\end{flalign}
The meaning of the two phases is obvious and corresponds to the discussion in the previous section. Note that the mean-field theory yields the same result that would be obtained by an effective Ising spin model. However, fluctuations around the mean-field solution are different in the original spin-boson lattice and thus, the validity of the mean-field description cannot be addressed with an effective Ising Hamiltonian.

\section{Gaussian fluctuations}
\label{sec:gaussian_fluctuations}

Quantum fluctuations are expected to diverge at the critical point. Near the latter our mean-field description is no longer valid. 
To quantify fluctuations we use a Gaussian approximation around the mean-field solution. As we are going to consider a particular case of the mean-field solution later on, which presents homogeneity in the spin states, we drop the index in $\theta_j$.

Let us define first fluctuations operators with respect to the bosonic degrees of freedom,  $\delta \bar{a}_n=\bar{a}_n-\bar{\alpha}_n$. 
Spin fluctuations are more involved and we quantify them by using Holstein-Primakoff bosons \cite{Auerbach}.
We have to work in a spin rotated frame such that the mean-field state is the reference state for the Holstein-Primakoff transformation. To this aim we define the rotated operators $\tilde{\sigma}_{j}^{x,y,z}$,
\begin{flalign}
&\left\{
\begin{array}{l}
\sigma_{j}^y=\tilde{\sigma}_{j}^y,\\
\\
\sigma_{j}^z=\cos \theta \tilde{\sigma}_{j}^z+\sin \theta \tilde{\sigma}_{j}^x,\\
\\
\sigma_{j}^x=\cos \theta \tilde{\sigma}_{j}^x-\sin \theta \tilde{\sigma}_{j}^z.
\end{array}
\right.
\end{flalign}
The rotated Hamiltonian (from the original Hamiltonian (\ref{Hamiltonian}) in the collective mode basis) becomes

\begin{flalign}
H_{\rm JT} &= \sum_{j}\frac{\Omega }{2} (\cos \theta \tilde{\sigma}_{j}^x-\sin \theta \tilde{\sigma}_{j}^z)+\sum_{n}\bar{\omega} _{n}\bar{a}_{n}^{\dagger}\bar{a}_{n}\notag\\
&+g\sum_{j,n}(\cos \theta \tilde{\sigma}_{j}^z+\sin \theta \tilde{\sigma}_{j}^x) (M_{j,n} \bar{a}_{n}+M_{j,n}^*\bar{a}_{n}^{\dagger }).
\end{flalign}
After the rotation, we can use the usual Holstein-Primakoff transformation \cite{Auerbach} where the reference state is taken in the $\tilde{\sigma}_{j}^z$ basis. The spin fluctuations are expressed by means of bosons $b_j$, $b_{j}^\dagger$,

\begin{flalign}
&\left\{
\begin{array}{l}
\sigma_{j}^+ = b_{j}^{\dagger} \sqrt{1 - n_{s,j}},\\
\\
\sigma_{j}^-=\sqrt{1 - n_{s,j}}\, b_j,\\
\\
\sigma_{j}^z = 2 n_{s,j} - 1,
\end{array}
\right.
\label{HP}
\end{flalign}
where $n_{s,j} = b_{j}^{\dagger} b_j$. It is straightforward to show that those operators indeed obey the commutation relations corresponding to the Pauli matrices.
In the limit of small fluctuations we approximate
\begin{flalign}
&\left\{
\begin{array}{l}
\sigma_{j}^+\simeq b_{j}^\dagger,\\
\\
\sigma_{j}^- \simeq b_{j}.\\
\end{array}
\right.
\end{flalign}
The former approximations are only valid in the limit $\langle b_{j}^\dagger b_j \rangle \ll 1$ which we shall verify later for self-consistency. Near the critical point, we expect that 
large spin fluctuations such that nonlinear terms in Eq. (\ref{HP}) become relevant.

Substituting in the rotated Hamiltonian we arrive at the Gaussian fluctuations Hamiltonian,
\begin{flalign}
&H_{\rm G}=\sum_n \bar{\omega}_n \delta \bar{a}_{n}^\dagger \delta \bar{a}_n + \sum_j \Delta_j b_{j}^\dagger b_{j}\notag\\
&+ \sum_{j,n} g \sin \theta (M_{j,n}^*\delta \bar{a}_{n}^\dagger+ M_{j,n}\delta \bar{a}_n)(b_{j}^\dagger+b_{j}),
\label{H.g}
\end{flalign}
where $\Delta_j:=-\Omega \sin \theta + 2 g \cos \theta \sum_n \left(M_{j,n} \bar{\alpha}_n+{\rm c.c.}\right)$. 
$H_\GG$ is diagonalized by means of a Bogoliubov transformation to spin-phonon fluctuation operators $c_{m}$,
\begin{equation}
H_{\rm G}=\sum_{m = 1, \dots, 2 N} E_m c_{m}^{\dagger} c_m.
\end{equation}
The new $2 N$ bosons are related to the $N$ boson and $N$ spin fluctuation operators by a relation of the form
\begin{eqnarray}
\delta \bar{a}_n &=&
\sum_{m = 1,\ldots, 2N} 
\left(
W_{n,m}^{(a)} c_m + V_{n,m}^{(a)} c_{m}^{\dagger} \right),
\notag \\
b_{j} &=&
\sum_{m = 1,\ldots, 2N} 
\left(
W_{j,m}^{(b)}c_{m} + V_{j,m}^{(b)} c_m^{\dagger}
\right),
\label{bogoliubov}
\end{eqnarray}
where the matrices $W_{n,m}^{(a)}, W_{j,m}^{(b)}, V_{n,m}^{(a)}, V_{j,m}^{(b)}$ define a generalized Bogoliubov  transformation, which has to satisfy the condition that bosonic commutation relations are left invariant. 
We will find below a very convenient way to get an explicit result for that transformation. For the moment, we just show how to compute quantum fluctuations. Define first the vacuum of the eigenmodes of $H_{\GG}$, $|\Omega \rangle$ by the condition $c_m |\Omega \rangle = 0$. Then we define the variance per atom for a set of the original spin-phonon fluctuation modes and get the result
\begin{eqnarray}
F_{\{ \delta a \}} &=& \frac{1}{N} \sum_n \langle \Omega | \delta \bar{a}_{n}^\dagger \delta \bar{a}_n |\Omega \rangle=\sum_{n,m} |V_{n,m}^{(a)}|^2,\notag\\
F_{\{ b\}} &=& \frac{1}{N} \sum_j \langle \Omega | b_{j}^\dagger b_{j} |\Omega \rangle=\sum_{j,m} |V_{j,m}^{(b)}|^2.
\label{fluctuations}
\end{eqnarray}
In the next section we use these expressions to calculate quantum fluctuations in the case of translational invariant systems with PBC.
\section{Analysis of the problem under Periodic Boundary Conditions}
\label{pbc_analysis}
To impose PBC we substitute the eigenfunctions $M_{j,n}$ defined in Eq. (\ref{pbc}). We transform fluctuation operators to the plane-wave basis by the relations 
$\delta a_j = \sum_n M_{j,n} \delta \bar{a}_n$, $b_j = \sum_n M_{j,n} \bar{b}_n$, and use the relations (\ref{alpha}, \ref{J}, \ref{critical}) in the Hamiltonian (\ref{H.g}), to get
\begin{eqnarray}
H_{\rm G} =
\sum_n \bar{\omega}_n \delta \bar{a}_n^{\dagger} \delta \bar{a}_n + 
\sum_n g \sin \theta (\delta \bar{a}_{-n}^{\dagger} + \delta \bar{a}_n)( \bar{b}_{n}^{\dagger}+  \bar{b}_{-n}) -\sum_n \frac{\Omega}{\sin \theta} \bar{b}_{n}^{\dagger} \bar{b}_{n}.
\end{eqnarray}
Note that the closure relation in the plane-wave basis yields couplings between modes with opposite linear momentum, $n$ and $-n$, in terms of the form $\delta \bar{a}_n \bar{b}_{-n}$, for example.
To diagonalize $H_\GG$ we work in the $X,P$ representation. We define
\begin{equation}
\left\{ 
\begin{array}{l}
X_{a,n}=\frac{1}{\sqrt{2\bar{\omega}_n }}(\delta \bar{a}_{-n}^{\dagger }+\delta \bar{a}_{n}), \\ 
P_{a,n}=\frac{1}{i}\sqrt{\frac{\bar{\omega}_n}{2}}(\delta \bar{a}_{-n}-\delta \bar{a}_{n}^{\dagger }),
\end{array}
~~~\left\{ 
\begin{array}{l}
X_{b,n}=\sqrt{\frac{|\sin\theta|}{2\Omega }}( \bar{b}_{-n}^{\dagger }+ \bar{b}_{n}), \\ 
P_{b,n}=\frac{1}{i}\sqrt{\frac{\Omega}{2|\sin\theta|}}( \bar{b}_{-n}- \bar{b}_{n}^{\dagger }).
\end{array}
\right. \right. 
\end{equation}
Note the relations
$X_{\mu,n}^\dagger = X_{\mu,-n}$, 
$P_{\mu,n}^\dagger = P_{\mu,-n}$ $(\mu = a,b)$. 
Up to an additive constant corresponding to the vacuum energy, the problem can be stated as a sum of separable Hamiltonians for each mode $n$,
\begin{eqnarray}
H_{\GG} &=& \sum_n H_{\GG,n},
\nonumber \\
H_{\GG,n} &=& 
\frac{1}{2} P_{a,n}P_{a,-n}+\frac{1}{2} \bar{\omega}_{n}^2 X_{a,n}X_{a,-n} +\frac{1}{2} P_{b,n}P_{b,-n}+\frac{1}{2} \left(\frac{\Omega}{|\sin\theta|}\right)^2 X_{b,n}X_{b,-n}
\nonumber \\
&-& g\sqrt{\Omega \bar{\omega}_n |\sin\theta|} \left( X_{a,n} X_{b,-n} + X_{b,n} X_{a,-n} \right).
\end{eqnarray}
In matrix notation each of the Hamiltonians $H_{\GG,n}$ can be written as  
\begin{flalign}
H_{\GG, n}
&=\frac{1}{2}\sum_{\mu,\nu} K_{\mu,\nu}^{(n)} X_{\mu,n} X_{\nu,-n} +\frac{1}{2} \sum_\mu P_{\mu,n} P_{\mu,-n}, \\&K^{(n)}=
\begin{pmatrix}
\bar{\omega} _{n}^{2} & -2g\sqrt{\Omega \bar{\omega} _{n}|\sin\theta|} \\ 
-2 g \sqrt{\Omega \bar{\omega} _{n}|\sin\theta|} & \left(\frac{\Omega}{|\sin\theta|}\right)^{2}
\end{pmatrix},\nonumber
\end{flalign}
where $\mu$ and  $\nu$ run over labels $a,b$.

Now we have to find a transformation defined by a matrix $U_{\mu,\nu}^{(n)}$, 
$\hat{X}_{\mu, n} = \sum_\nu U_{\mu,\nu}^{(n)} X_{\nu, n}$,
$\hat{X}_{\mu,-n} = \sum_\nu U_{\mu,\nu}^{(n)} X_{\nu,-n}$,
$\hat{P}_{\mu, n} = \sum_\nu U_{\mu,\nu}^{(n)} P_{\nu,n}$,
$\hat{P}_{\mu,-n} = \sum_\nu U_{\mu,\nu}^{(n)} P_{\nu,-n}$ that diagonalizes 
$K^{(n)}$,
and conserves the canonical commutation relations. The latter requirement is fulfilled automatically since the transformation is orthogonal due to the fact that $K^{(n)}$ is real and symmetric.

\begin{figure}
\resizebox{0.9\columnwidth}{!}{
\includegraphics{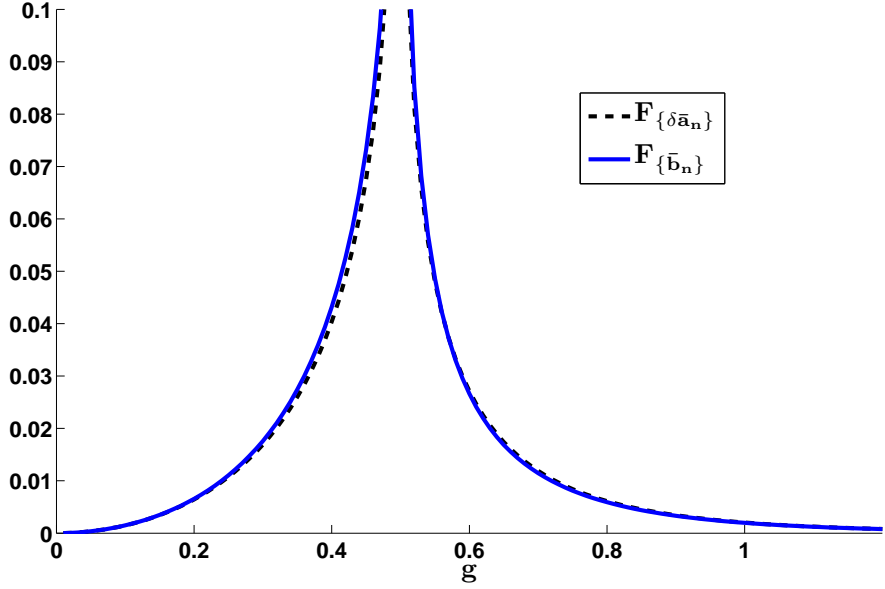}}
\caption{Quantum Gaussian fluctuations for $N=20$ ions. Black and blue lines correspond to phonons and spin-waves respectively. We have considered homogeneous chains and units such that $\Omega=\bar{\omega}_0=1$, $t = 0.4$. The critical point is at $g_{\rm c} = 1/2$.}
\label{fluc1}
\end{figure}

The eigenvalues of $K^{(n)}$ can be split up into two branches,
\begin{equation}
E_{\pm,n}^2=\frac{\left(\frac{\Omega}{|\sin\theta|}\right)^2+\bar{\omega}_{n}^2 \pm \sqrt{16g^2\Omega \bar{\omega}_n|\sin\theta| + \left(\left(\frac{\Omega}{|\sin\theta|}\right)^2-\bar{\omega}_{n}^2\right)^2}}{2}.
\end{equation}
The transformed eigenvectors $\hat{X}_{\pm,n}$ and $\hat{P}_{\pm,n}$ are the columns of the $U^{(n)}$ matrix
\begin{equation}
U^{(n)}=\begin{pmatrix}
-\frac{2g\sqrt{\Omega \bar{\omega}_n|\sin\theta|}}{v_n} & \frac{E_{-,n}^2-\left(\frac{\Omega}{|\sin\theta|}\right)^2}{v_n} \\
\frac{E_{+,n}^2-\bar{\omega}_{n}^2}{v_n} & -\frac{2g\sqrt{\Omega \bar{\omega}_n|\sin\theta|}}{v_n}
\end{pmatrix};
\end{equation}
there the normalization factor is
\begin{equation}
v_n^2 = 
\left(E_{-,n}^2-\left(\frac{\Omega}{|\sin\theta|}\right)^2\right)^2+4g^2 \Omega \bar{\omega}_{n}|\sin\theta|.
\end{equation}
We write the new operators $\hat{X}_{\pm,n}$,$\hat{P}_{\pm,n}$ in second quantized form,
\begin{equation}
\left\{ 
\begin{array}{l}
\hat{X}_{\pm,n}=\frac{1}{\sqrt{2 E_{\pm,n} }}(c_{\pm,-n}^{\dagger}+c_{\pm,n}), \\ 
\hat{P}_{\pm,n}=\frac{1}{i}\sqrt{\frac{E_{\pm,n}}{2}}(c_{\pm,-n}-c_{\pm,n}^{\dagger}).
\end{array}
\right.
\end{equation}
From here we can easily get a transformation to relate 
$\delta \bar{a}_{s, n}$, $\delta \bar{a}_n$ 
to normal modes $c_{\pm,n}$ like in Eq. (\ref{bogoliubov}) 
and use those to get a final expression for the fluctuations defined in 
Eq. (\ref{fluctuations}),
\begin{eqnarray}
F_{\{\delta a\}} &=& \sum_n F_{\{\delta \bar{a}_n\}},
\notag \\
F_{\{ \delta \bar{a}_n \}} &=&
\frac{1}{N} \frac{1}{v_{n}^2}  
\left(
g^2\Omega |\sin\theta| E_{+,n} \left(\frac{\bar{\omega}_n}{E_{+,n}}-1\right)^2
\right.
\notag \\
&& \hspace{2cm}
+
\left.
\frac{\left(E_{-,n}^2-\frac{\Omega^2}{|\sin\theta|^2}\right)^2}{4} \frac{E_{-,n}}{\bar{\omega}_n} \left(1-\frac{\bar{\omega}_n}{E_{-,n}}\right)^2 
\right), 
\notag \\
F_{\{b\}} &=& \sum_n F_{\{\bar{b}_{n}\}},
\notag \\
F_{\{ \bar{b}_{n}\}} &=& \frac{1}{N} \frac{1}{v_{n}^2}
 \left( g^2\bar{\omega}_n |\sin\theta|^2 E_{-,n} \left(\frac{\Omega}{|\sin\theta|}\frac{1}{E_{-,n}}-1\right)^2 \right.
\notag \\
&& \hspace{2cm} 
+ 
\left. \frac{(E_{+,n}^2-\bar{\omega}_{n}^2)^2}{4} \frac{1}{E_{+,n}}\frac{\Omega}{|\sin\theta|} \left(1-\frac{E_{+,n}}{\Omega}|\sin\theta|\right)^2 \right)
\label{fluctuations.pbc}
\end{eqnarray}
In the following we are going to analyze the results obtained by means of Eq. (\ref{fluctuations.pbc}).
In Fig. \ref{fluc1} we plot the quantum fluctuations as a function of $g$ and check that they diverge at the critical point. Although this divergence is a straightforward effect it will shed light on the scaling of fluctuations with $N$. We have chosen units such that the spin and boson fluctuations are of the same order.
The mean-field critical point is defined by the condition 
$\Omega = 2 J = 4 g^2/\bar{\omega}_0$. 

First we notice that $E_{-,n} = 0$ if $n=0$ since \smash{$E_{-,n}\overset{n\rightarrow 0}{\longrightarrow} c\cdot n$}, with $c$ a constant determining the speed of acoustic fluctuations at the critical point. Therefore from (\ref{fluctuations.pbc}) it can be shown that the the fluctuations are dominated by the term $1/E_{-,n}$. This fact, indeed, makes the fluctuations ill-defined for $n=0$.
On the other hand the contributions $n\neq 0$ to $1/E_{-,n}$ in (\ref{fluctuations.pbc}) are the source of a logarithmic divergence expected as 
$N\rightarrow \infty$,  which can be studied separately from the $n=0$ divergence, by looking at the well-defined modes with $n \neq 0$,
\begin{equation}
\frac{1}{N} \sum_{n=1}^N \frac{1}{E_{-,n}}=\frac{1}{N} \sum_{n=1}^{N} \frac{1}{E \left(\frac{n}{N} \right)} \sim \frac{1}{N} \int_1^N  \frac{dn}{E \left(\frac{n}{N} \right)}= \int_{1/N}^1 \frac{dx}{E(x)} \sim \log N.
\label{log.div}
\end{equation}
We conclude that the amplitude of quantum fluctuations has to be understood by studying separately two different contributions:
\begin{itemize}
\item Modes $n=0$ show the same behaviour as the one expected for the Dicke model: fluctuations diverge at the critical point, however, they are suppressed for $g \neq g_{\rm c}$. As $N \to \infty$, the critical point becomes singular and any $g \neq g_c$ shows no fluctuations. 
\item Modes $n \neq 0$ contribute with a logarithmic divergence at the critical point due to the infrared divergence in the fluctuations. Out of the critical point, we expect an infrared cut-off in the fluctuation spectrum, and thus there is a well-defined $N \to \infty$ limit. 
\end{itemize}
Out of this discussion we get the following picture: Exactly at the critical point ($g = g_{\rm c}$) fluctuations diverge and they are governed by the $n=0$ mode. Out of the critical point both $n = 0$ and $n \neq 0$ modes contribute but the $n = 0$ contribution will be suppressed in the thermodynamic $N \to \infty$ limit. 

\begin{figure}
\resizebox{0.9\columnwidth}{!}{
\includegraphics{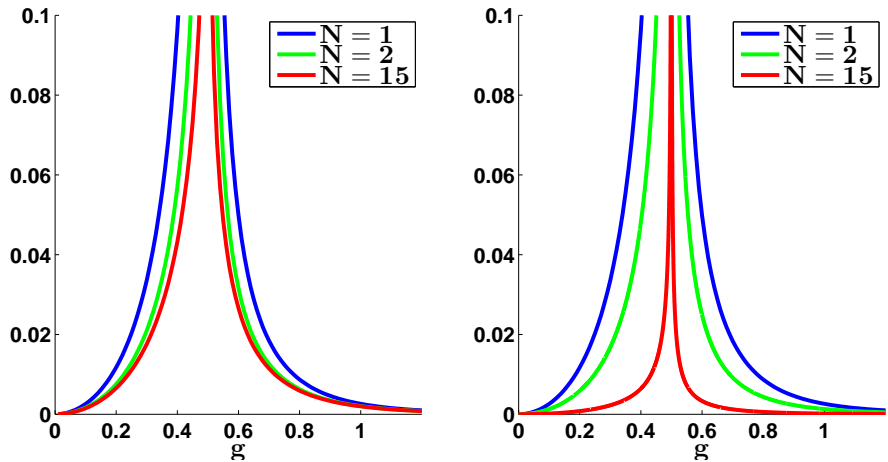}}
\caption{Quantum Gaussian fluctuations for increasing number of ions $N$ including all the Bogoliubov modes $n$ (left) and considering just the $n=0$ mode (right). 
We choose values $\bar{\omega}_0 = 1$, $\Omega = 1$, $t = 0.4$.
We depict only the spin-waves fluctuations $F_{\{ \bar{b}_{n}\}}$.} 
\label{fluc2}
\end{figure}

\begin{figure}
\resizebox{0.9\columnwidth}{!}{
\includegraphics{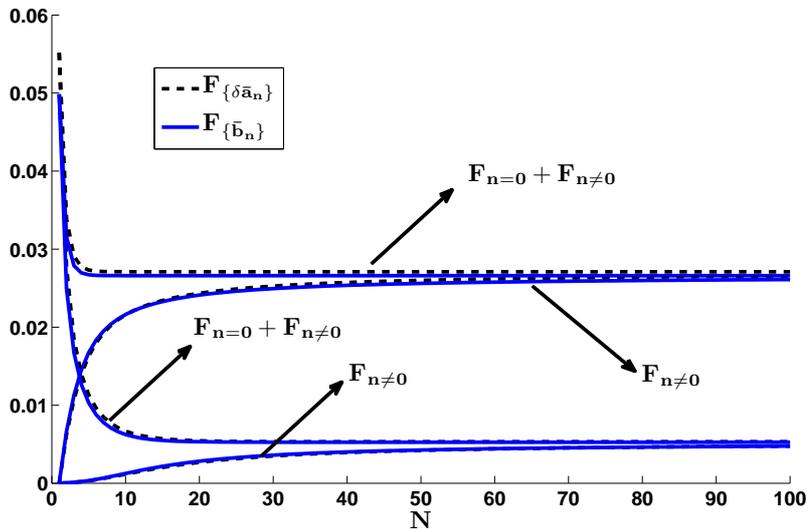}}
\caption{Quantum Gaussian fluctuations for values $\Omega=\bar{\omega}_0=1$ and $g=0.6$. We have chosen two values for the hopping: $t=0.4$ (up) and $t=10$ (down).}
\label{FvsN}
\end{figure}

\begin{figure}
\resizebox{0.9\columnwidth}{!}{
\includegraphics{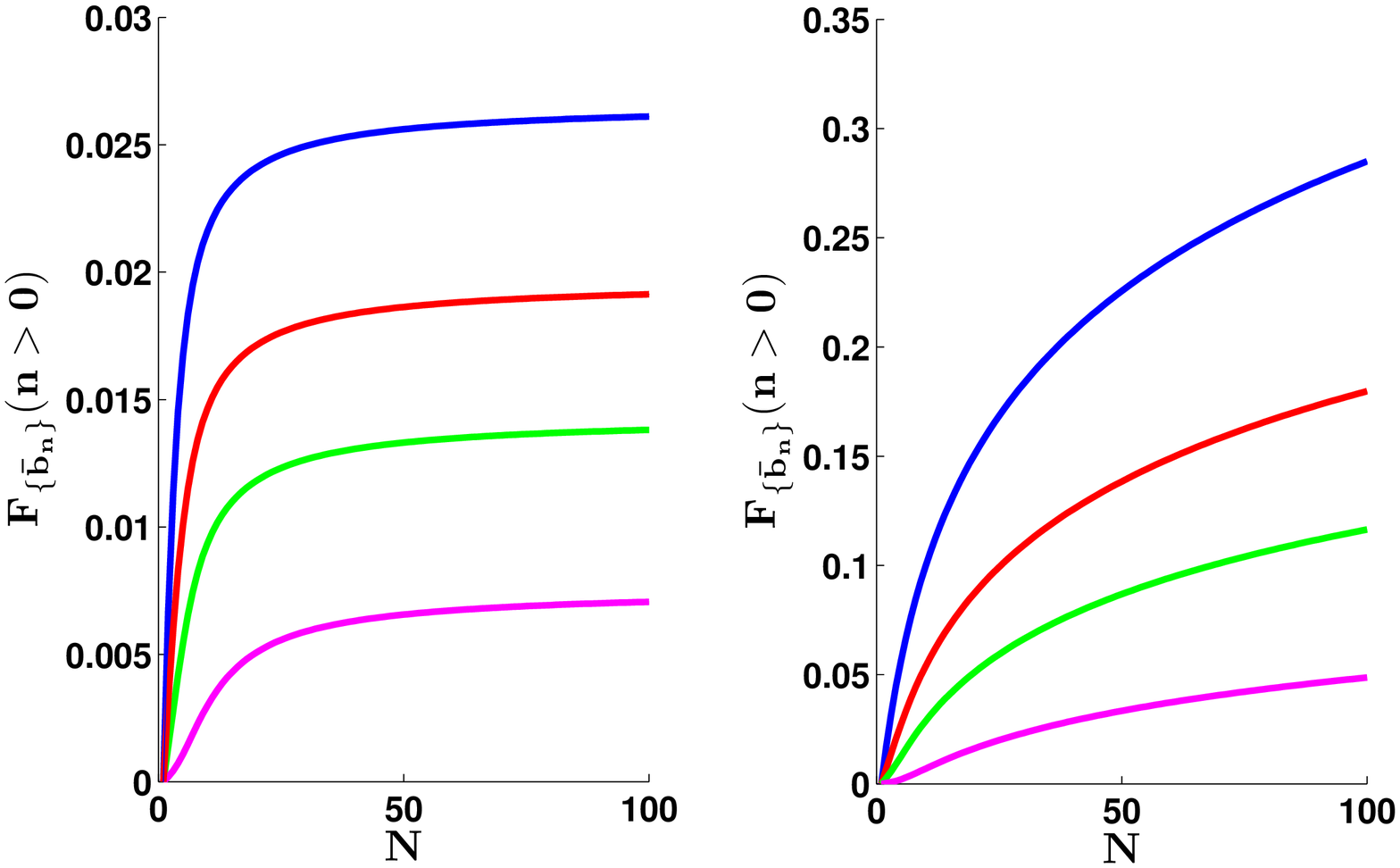}}
\caption{Spin-waves fluctuations for $t=0.4,0.8,1.5,5$ (from top to bottom) with $\Omega=\bar{\omega}_0=1$ and $g=0.6$ (left), $g=g_c=0.5$ (right). For increasing $t$ fluctuations are lower.}
\label{FvsT}
\end{figure}

Following those arguments, we calculate separately the total fluctuations 
($F_{b}$, $F_{\delta a}$), 
the contribution from $n = 0$, 
($F_{\bar{b}_{0}}$, $F_{\delta \bar{a}_0}$), 
and the fluctuations from modes $n \neq 0$,  
($\sum_{n \neq 0} F_{\bar{b}_{n}}$, $\sum_{n \neq 0} F_{\delta \bar{a}_n}$). 
In Fig. \ref{fluc2} (left), we plot the total fluctuations as function of $g$ for different values of $N$. Note that, for concreteness, we consider the case $\Omega = \omega_0$, which leads to similar spin and boson quantum fluctuations.
We find convergence in the large $N$ limit. In Fig. \ref{fluc2} (right) we plot the $n=0$ contribution and check that it contributes to the divergence at the critical point, but it can be neglected in the noncritical region for large $N$, as expected from the single-mode Dicke model.
In Fig. \ref{FvsN} we plot the $n\neq0$ contribution and the total fluctuations as a function of the number of particles. We observe that fluctuations converge to the $n \neq 0$ contribution for large $N$, however, for the values used in this calculations there is an intermediate {\rm mesoscopic} regime of $N \approx 10$ particles, where fluctuations decrease with $N$. This effect is induced by the suppression of the $n=0$ contribution as $N$ increases. Finally, Fig. \ref{FvsT} shows the scaling of the $n \neq 0$ contribution for different values of $t$. 
At the critical point (right), we check the logarithmic divergence estimated in Eq. 
(\ref{log.div}). Out of the critical phase, fluctuations converge to a steady value. A crucial observation is that for increasing $t$, the contribution from $n \neq 0$ decreases, due to the energy cost of $n \neq 0$ fluctuations.
\section{Conclusions and Outlook}
We have presented a mean-field theory for cooperative Jahn-Teller models that appear in a natural way in a variety of quantum optical setups. Our investigation has relied on the use of mean-field and spin-wave theory. The latter is limited in what concerns the description of the critical phase. However, we have shown that there is a regime of validity for mean-field theory that can be checked self-consistently by calculating the amplitude of quantum fluctuations. Our results show that the mean-field phase is determined by the interplay between the $n=0$ fluctuations typical from long-range Dicke models and the $n \neq 0$ fluctuations that are well-defined in the thermodynamic limit. Our calculations show that quantum fluctuations decrease with $N$ due to the suppression of the $n=0$ contribution and arrive at a steady value for large values of $N$.

Our work is relevant to several experimental setups in quantum optics. In particular, in the case of trapped ions, it was recently shown that a generalization of the cooperative Jahn-Teller model (\ref{Hamiltonian}) could be implemented. This can be achieved either inducing interactions with lasers \cite{Porras04aprl} or with magnetic field gradients \cite{Porras12bprl}. Typical energy scales are $\bar{\omega}_0, \Omega, g, t \approx 100$ kHz,
and ion chains of $N = 2, \dots 50$ ions. Another promising system is circuit QED, where qubit-field couplings in the ultrastrong coupling correspond to values $g \approx \bar{\omega}_0, \Omega$ in the GHz regime \cite{Schoelkopf08nat,Houck12natphys}. Recent theoretical proposal could allow one to induce sidebands in the qubit-field coupling to achieve a high degree of controllability of the parameters of the model \cite{Porras12aprl}.
In addition, our work could be complemented by the study of Jaynes-Cummings-like qubit-field interactions, which also arise in a variety of quantum optical systems.

An interesting issue that could be addressed by means of spin-wave theory is the effect of finite temperature. Experimental quantum optical systems like trapped ions and circuit QED systems are typically out-of-equilibrium systems. Ground states are created by adiabatic evolution, so that the correct physical description would correspond to an initial finite temperature state which evolves by adiabatically turning on some Hamiltonian parameter (for example, the spin-boson coupling, g). The theoretical description would thus rely on a non-equilibrium spin-wave theory to account for temporal evolution. Finite temperature effects would then lead to a finite number of gaussian excitations in the final state after the adiabatic evolution. 

We acknowledge QUITEMAD S2009-ESP-1594, FIS2009-10061, CAM-UCM/910758 and RyC Contract Y200200074, and COST action "IOTA".

\end{document}